**Tau creates spatially moving hotspots of amyloid-β secondary nucleation: a reaction–diffusion model**

**Andrey V. Kuznetsov**

Department of Mechanical and Aerospace Engineering, North Carolina State University, Raleigh, NC 27695-7910, USA; Email: avkuznet@ncsu.edu

## Abstract

Tau–amyloid-β cross-seeding creates a spatial paradox. Bioavailable tau can promote Aβ42 nucleation, thereby increasing local fibril number and catalytic-site burden; yet the same fibrils consume the free Aβ monomer required for homotypic secondary nucleation. The region containing the most tau or fibrillar material therefore need not be the region with the highest secondary-nucleation rate. We test this idea with a reaction–diffusion model that tracks a propagating effective tau catalyst, free Aβ monomer, elongation-competent fibrils, fibril mass, and the catalytic sites exposed to monomer. The model reduces the spatial competition to two quantities: accessible catalytic sites and free monomer. Their balance yields an exact condition for an interior kinetic maximum: the logarithmic decrease in accessible catalytic sites must equal twice the logarithmic increase in free monomer. Two Damköhler numbers then separate the relevant competitions—elongation versus clearance, and elongation versus diffusive redistribution—while a monomer screening length sets the physical scale over which depletion can persist. Independent transport and turnover estimates place this length near 0.44–0.54 mm. Simulations show that tau first raises secondary nucleation near its source; continued fibril growth then produces a monomer-depleted wake, and the maximum shifts to a region with fewer catalytic sites but more available monomer. Oligomer transport and sustained reversible exchange can weaken or eliminate this contrast, whereas growth-history-dependent site density can strengthen it. The robust prediction is therefore a spatial separation among tau burden, fibril-mass growth, and instantaneous Aβ secondary-nucleation activity, rather than a universal hotspot speed.

## 1. Introduction

The central question of this paper is simple: as tau pathology spreads through tissue, where should Aβ secondary nucleation be most active? The answer need not be the region with the greatest tau burden. Aβ plaques are extracellular, tau aggregates originate intracellularly, and the two pathologies evolve on different spatial and temporal scales. Existing spreading models describe where pathological burden moves, whereas molecular aggregation models describe how nucleation, elongation, fragmentation, and clearance change local protein populations. Here we connect those two descriptions by asking where the instantaneous Aβ secondary-nucleation rate becomes maximal during the spread of a tau perturbation.

Two recent experimental results make this question concrete. In one study (Mosconi et al. 2026), disease-relevant paired-helical-filament (PHF) and chronic-traumatic-encephalopathy (CTE) tau fibrils catalyzed Aβ42 primary nucleation in a fold-dependent manner. Tau can therefore increase the local number of growing Aβ fibrils. In independent work (Hu et al. 2026), Aβ secondary nucleation was found to be concentrated at rare catalytic defects rather than being uniformly proportional to fibril surface. These findings imply that four quantities that are often conflated—fibril mass, fibril number, catalytic-site abundance, and free monomer concentration—must be treated separately.

A companion active-site-resolved model made this separation in a spatially lumped tau–Aβ system (Kuznetsov 2026a). It showed that tau-induced fibril shortening does not by itself make Aβ fibrils more catalytic at matched mass. It also identified a kinetic switch governed by the Damköhler number $\mathrm{Da} = 2k_+P/\lambda_m$, denoted $\mathrm{Da}_{\mathrm{EC}}$ below. Here $k_+$ is the fibril elongation coefficient, $P$ is the concentration of elongation-competent fibrils, the factor 2 counts the two growing fibril ends, and $\lambda_m$ is the first-order monomer-clearance coefficient. Thus $\mathrm{Da}_{\mathrm{EC}}$ compares monomer consumption by elongation with monomer removal by clearance. Section 3.3 restates the same ratio in the reduced variables of the spatial model. When $\mathrm{Da}_{\mathrm{EC}}$ is small, clearance controls the free-monomer pool; when $\mathrm{Da}_{\mathrm{EC}}$ is large, elongation becomes a strong monomer sink.

Introducing space turns that kinetic switch into a new mechanism. A region newly reached by tau may already contain additional fibrils and catalytic sites while still retaining much of its free Aβ monomer. A region exposed to tau for longer may contain even more fibrillar material, but it can also contain many more elongating fibril ends and therefore much less free monomer. The first region has fewer catalytic sites but more substrate; the second has more sites but less substrate. Because homotypic secondary nucleation requires both, its maximum can lie between them and can move as the tau perturbation advances.

Throughout this paper, *hotspot* means only a local maximum of the instantaneous homotypic Aβ secondary-nucleation rate. This kinetic definition should not be confused with the protofibrillar Aβ42 hotspots observed around plaques with incomplete microglial coverage (Condello et al. 2015), which reflect additional local biology. The question here is narrower: can transport and aggregation kinetics alone create a spatially displaced, time-dependent maximum of secondary nucleation?

The paper is organized around that question. Section 2 builds the model in four steps: a moving tau field, local Aβ aggregation kinetics, a closure for catalytic-site accessibility, and the reduced four-field system used in the simulations. Every state variable and coefficient is defined when it first appears; Table 1 then summarizes numerical values rather than serving as a glossary. Section 3 derives the condition that locates an interior hotspot and introduces the two Damköhler numbers. Section 4 states the numerical method. Section 5 follows the mechanism from its formation to its physical scaling and then tests processes that can erase or reinforce it. The analysis is conditional on a positive tau→Aβ heterotypic coupling, which is well supported by the motivating *in vitro* data but remains less certain *in vivo*.

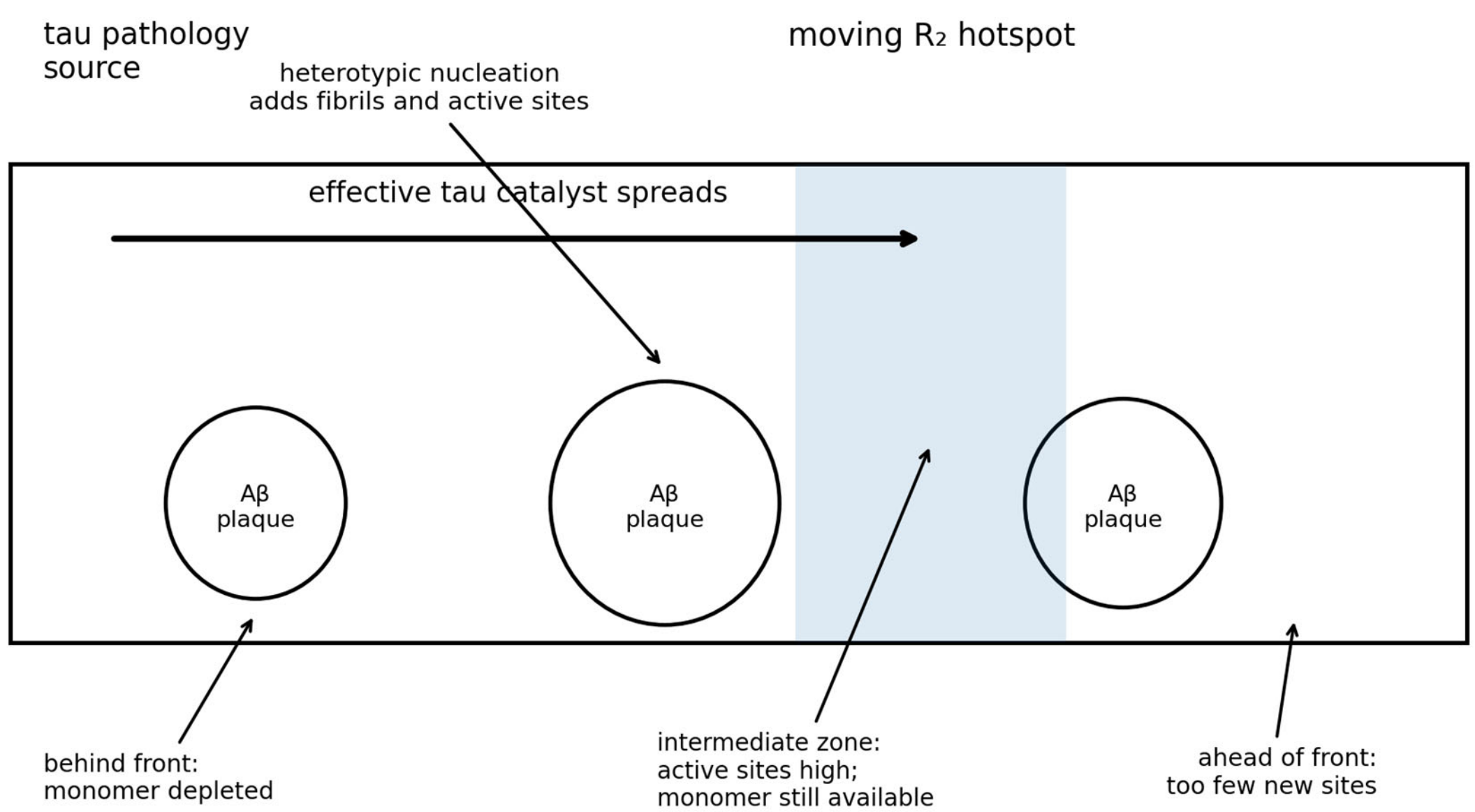

**Fig. 1. Conceptual structure of the spatial model. An effective bioavailable tau catalyst spreads from an initially localized source. Tau-driven heterotypic Aβ nucleation increases local fibril number and catalytic-site burden. Behind the advancing perturbation, accumulated fibril ends deplete free monomer; far ahead of it, too few new active sites have been generated. The homotypic secondary-nucleation maximum can therefore lie between the depleted wake and the weakly exposed region. In the Fisher–Kolmogorov implementation it may also lie ahead of a chosen tau-concentration contour because the tau field has a diffuse leading tail; "ahead of the tau front" below always means ahead of the stated contour, not ahead of all bioavailable tau.**

## 2. Mathematical model

How to read the model. The paper uses two levels of description. Eqs. (1)–(10) retain the physical variables and coefficients needed to connect the spatial model to the earlier biochemical formulation. Eqs. (11)–(15) are the reduced, dimensionless four-field system actually used for the reference simulations. The same physical processes appear at both levels, but several coefficients are rescaled or combined.

### 2.1. Step 1: represent bioavailable tau as a moving spatial field

The first step is to replace the companion model's single lumped tau-accessibility factor (Kuznetsov 2026a) with a spatial field. We denote by $T(x,t)$ the local amount of tau fibrillar material that is effectively available to catalyze Aβ nucleation at position $x$ and time $t$. $T$ is a coarse-grained catalyst field, not the concentration of a single molecular tau species: release, exposure, transport, uptake/re-release, and local amplification are all absorbed into it at the chosen tissue scale.

We describe the propagation and local amplification of this effective catalyst with a Fisher–Kolmogorov equation,

$$\frac{\partial T}{\partial t} = D_T \frac{\partial^2 T}{\partial x^2} + r_T T(1 - T/T_{\max}) \tag{1}$$

In Eq. (1), $D_T$ is the effective tau-catalyst spreading coefficient, $r_T$ is the local amplification rate, and $T_{\max}$ is the saturation level of $T$. The diffusion-like term represents net spatial spreading after unresolved transport processes are coarse-grained; the logistic term represents local amplification and saturation. This is therefore a mesoscale closure, not a claim that tau fibrils diffuse freely through homogeneous tissue.

The Fisher–Kolmogorov closure is used only where a moving perturbation is meaningful. A regional analysis (Meisl et al. 2021) found that late tau accumulation becomes increasingly replication-limited, whereas early and focal pathology remains spatially inhomogeneous. We therefore use Eq. (1) to represent early/focal propagation and within-region spread, not late neocortical progression. As an order-of-magnitude check, combining the 0.6 $\text{month}^{-1}$ replication rate reported for P301S mice (Meisl et al. 2021) with an estimated 0.6–1.2 mm per month spread from the semiquantitative pathology maps (He et al. 2018) gives $D_T \approx 6 \times 10^{-14}$ to $2 \times 10^{-13}$ $\text{m}^2\ \text{s}^{-1}$ through $c = 2(D_T\, r_T)^{1/2}$. This estimate is used only to set the physical scale later in the paper.

### 2.2. Step 2: describe local Aβ aggregation and active-site kinetics

At each position, the full Aβ subsystem distinguishes five local state variables: $m(x,t)$, the free Aβ monomer concentration; $S(x,t)$, soluble on-pathway oligomer equivalents; $P(x,t)$, the concentration of elongation-competent fibrils; $M(x,t)$, fibril mass; and $D(x,t)$, the concentration of active secondary-nucleation sites. Keeping these variables separate is essential. Tau-driven nucleation can increase $P$, $M$, and $D$, while the additional fibril ends simultaneously consume $m$. Intrinsic primary, tau-catalyzed heterotypic, and homotypic secondary nucleation occur at the local rates

$$R_1 = k_{o1} m^2 \tag{2}$$

$$R_\tau = k_\tau T m^2 \tag{3}$$

$$R_2 = k_{o2,D} D_{\text{eff}} m^2 \tag{4}$$

The three nucleation rates in Eqs. (2)–(4) are defined as follows. $R_1$ is intrinsic Aβ primary nucleation, $R_\tau$ is tau-catalyzed heterotypic nucleation, and $R_2$ is homotypic Aβ secondary nucleation. The corresponding rate coefficients are $k_{o1}$, $k_\tau$, and $k_{o2,D}$, respectively. The field $T$ is the effective tau catalyst defined in Eq. (1), $m$ is free monomer, and $D_{\text{eff}}$ is the portion of the active-site population $D$ that remains accessible to soluble Aβ. The common $m^2$ factor reflects the experimentally fitted second-order dependence discussed below. The complete local mass and population balances are then

$$\frac{\partial m}{\partial t} = D_m \frac{\partial^2 m}{\partial x^2} + q - R_1 - R_\tau - R_2 + R_{\text{diss}} - 2k_+ mP - k_{\text{conv}} mS - \lambda_m m \tag{5}$$

$$\frac{\partial S}{\partial t} = D_S \frac{\partial^2 S}{\partial x^2} + R_1 + R_\tau + R_2 - R_{\text{diss}} - k_{\text{conv}} mS - \lambda_S S \tag{6}$$

$$\frac{\partial P}{\partial t} = k_{\text{conv}} mS + k_{\text{frag}} M - \lambda_F P \tag{7}$$

$$\frac{\partial M}{\partial t} = 2k_+ mP + 2k_{\text{conv}} mS - \lambda_F M \tag{8}$$

$$\frac{\partial D}{\partial t} = \eta_{\text{birth}} k_{\text{conv}} mS + 2\eta_{\text{grow}} k_+ mP + \eta_{\text{frag}} k_{\text{frag}} M - (k_{\text{mask}} + \lambda_F) D \tag{9}$$

Eqs. (5)–(9) introduce the remaining dimensional coefficients. $D_m$ and $D_S$ are the spatial diffusion coefficients of monomer and soluble oligomer equivalents, respectively; $q$ is the local monomer-production rate; and $R_{\text{diss}}$ is the rate at which soluble oligomeric material dissociates back toward monomer. The coefficient $k_+$ is the fibril-elongation rate constant, so $2k_+ mP$ is monomer consumption by the two fibril ends. The coefficient $k_{\text{conv}}$ is the oligomer-to-fibril conversion coefficient; $\lambda_m$ and $\lambda_S$ are first-order removal coefficients for monomer and soluble oligomers; $k_{\text{frag}}$ is the fibril-fragmentation coefficient; and $\lambda_F$ is the fibril-removal coefficient. In the active-site balance, $\eta_{\text{birth}}$, $\eta_{\text{grow}}$, and $\eta_{\text{frag}}$ are the numbers of active sites created per fibril birth, per unit elongational growth, and per fragmentation event, while $k_{\text{mask}}$ is the site-masking rate. Thus Eqs. (5)–(9) provide the full biochemical bookkeeping from which the reduced spatial model is derived.

The quadratic dependence on free monomer is important because it gives monomer recovery strong leverage over hotspot position. Kinetic fits (Mosconi et al. 2026) treated the secondary-nucleation reaction order as a free parameter and yielded values close to 2 before the order was fixed at 2; the primary-nucleation critical nucleus size was likewise set to 2 based on fits of Aβ42 self-aggregation. The same fits require saturation at much higher monomer concentrations: half-saturation occurs at approximately 36–449 nM, whereas diffusible extracellular Aβ42 in the frontal cortex of CRND8 mice is only about 160–425 pM (Waters 2010). At those tissue concentrations, the saturated secondary-nucleation law differs negligibly from the quadratic form. The factor of 2 in the hotspot condition derived later therefore reflects the low-concentration kinetics rather than a convenient mathematical choice.

### 2.3. Step 3: convert fibril mass into accessible catalytic sites

Secondary nucleation depends on sites that soluble Aβ can reach, not simply on the total fibril mass stored inside a deposit. We write the total active-site concentration in the reference limit as $D = \sigma_0 M$, where $\sigma_0$ is the number of active sites per unit fibril mass. Plaque compaction can nevertheless hide an increasing fraction of those sites. To represent that geometric loss of accessibility smoothly, we define an accessible-capacity function $A(M)$,

$$A(M) = \frac{M}{1 + (M/M_c)^{1/3}} \tag{10}$$

In Eq. (10), $A(M)$ is a dimensionless measure of the catalytic sites exposed to monomer, $M$ is fibril mass, and $M_c$ is the fibril-mass scale at which shielding becomes important. For $M \ll M_c$, $A(M) \approx M$ and accessibility is approximately proportional to mass. For $M \gg M_c$, $A(M) \propto M^{2/3}$, so exposed capacity grows like a surface area rather than a volume. We therefore take $D_{\mathrm{eff}}$ to be proportional to $A(M)$ in the reduced model.

The evidence for this closure is indirect. Measurements (Hu et al. 2026) showed that the number of catalytic sites was proportional to fibril mass within a given population of fibrils in solution, with approximately one Brichos binding site per 143 Aβ42 monomers; there is no molecular plaque-shielding term in those measurements. By contrast, longitudinal plaque imaging is consistent with surface-limited growth: newly formed and pre-existing plaques grow radially at similar rates in APPPS1 mice (Hefendehl et al. 2011), and growth slows at high plaque density in Tg2576 mice (Burgold et al. 2014). Eq. (10) should therefore be read as a tissue-scale accessibility closure, not as an experimentally measured law for catalytic-site exposure.

The reference calculations also keep $\sigma_0$ constant. That assumption is convenient but not fundamental. A study (Hu et al. 2026) showed that, for Aβ40, fibrils formed slowly near equilibrium carried substantially fewer catalytic defects than more rapidly grown fibrils, despite no corresponding morphological difference. Whether the same growth-history dependence applies to tau-seeded Aβ42 is unknown. Section 5.8 therefore uses the Aβ40 result only as a sensitivity scale, not as a calibrated Aβ42 parameter.

### 2.4. Step 4: reduce the full kinetics to the four-field spatial model actually solved

The full biochemical system contains more detail than is needed to isolate the spatial mechanism. The simulations therefore use four state fields: $T$, the effective tau catalyst; $m$, free Aβ monomer; $P$, elongation-competent fibrils; and $M$, fibril mass. The soluble-oligomer step is collapsed into an effective fibril-birth term, and the uniformly distributed intrinsic-primary-nucleation background is subtracted because it changes the baseline but cannot create a moving spatial maximum. Concentrations are scaled by reference values, $x$ by a mesoscopic length $L$, and time by a reference time scale. The resulting dimensionless system is

$$\frac{\partial T}{\partial t} = D_T \frac{\partial^2 T}{\partial x^2} + r_T T(1 - T) \tag{11}$$

$$\frac{\partial m}{\partial t} = D_m \frac{\partial^2 m}{\partial x^2} + \lambda(1 - m) - 2k_E m P - \nu(R_\tau + R_2) \tag{12}$$

$$\frac{\partial P}{\partial t} = R_\tau + R_2 - \mu_P P \tag{13}$$

$$\frac{\partial M}{\partial t} = 2k_E m P + \nu(R_\tau + R_2) - \mu_M M \tag{14}$$

$$R_\tau = k_\tau T m^2, \qquad R_2 = k_2 A(M) m^2 \tag{15}$$

All symbols in Eqs. (11)–(15) are dimensionless reduced quantities. $D_T$ and $D_m$ are the reduced tau and monomer redistribution coefficients; $r_T$ is the reduced tau-amplification rate; $\lambda$ is the reduced monomer restoration/clearance coefficient; $k_E$ is the reduced elongation coefficient; $\mu_P$ and $\mu_M$ are effective loss coefficients for fibril number and fibril mass; and $v$ is the fibrillar mass assigned per nucleation event. The two nucleation terms are $R_\tau = k_\tau T_{m2}$ and $R_2 = k_2 A(M) m_2$, where $k_\tau$ is the reduced heterotypic tau–Aβ nucleation coefficient and $k_2$ is the reduced homotypic secondary-nucleation coefficient. The state variables $T$, $m$, $P$, and $M$ retain the meanings given above, but are now scaled. In relation to the full model, $k_E$ plays the role of the scaled $k_+$, $\lambda$ the role of scaled $\lambda_m$, and $k_2$ contains the scaled secondary-nucleation activity associated with $k_{o2,D}$ together with the proportionality between $D_{\text{eff}}$ and $A(M)$. Thus no new physical process is introduced in the reduction; only fast or spatially uniform detail is removed. Table 1 lists the numerical values used below: a reference set chosen to make the mechanism visible and a mapped set derived from measured transport and turnover data; Section 2.5 explains how each value was obtained.

**Table 1. Numerical values used in the reduced model. The reference column contains exploratory dimensionless values used to demonstrate the mechanism in Sections 5.1–5.4; the mapped column contains values used for the physically interpreted calculations in Sections 5.5–5.8. Symbols are defined in the text before this table. One time unit is 0.50 months and one length unit is $L$ = 7.20 mm; on longer segments, mapped $D_T$ and $D_m$ are rescaled by (7.20 mm/$L$)$^2$ so that the corresponding physical transport coefficients remain unchanged.**

| Parameter | Meaning | Reference value | Mapped value | Reference or estimation method | Role in the calculation |
|---|---|---|---|---|---|
| $D_T$ | effective tau-catalyst spreading coefficient | 0.0025 | 0.0025 | With $r_T$, fixes the length unit: matching to 1 × $10^{-13}$ $m^2$ $s^{-1}$, within the 6 × $10^{-14}$ to 2 × $10^{-13}$ $m^2$ $s^{-1}$ obtained in Section 2.1 by combining a replication rate (Meisl et al. 2021) with an estimated spreading speed (He et al. 2018), gives $L$ = 7.20 mm (Sections 2.1, 2.6). The $T$ = 0.1 contour then advances at 0.79 mm per month, within the 0.6–1.2 mm per month range estimated from (He et al. 2018). | sets tau-front width and speed |

| Parameter | Meaning | Reference value | Mapped value | Reference or estimation method | Role in the calculation |
|---|---|---|---|---|---|
| $r_T$ | local tau-catalyst amplification rate | 0.30 | 0.30 | Fixes the time unit: matching to the tau-seed replication rate of 0.6 month$^{-1}$ in P301S mice (Meisl et al. 2021) gives 0.50 months (Section 2.6). | sets tau amplification and front speed |
| $D_m$ | effective monomer redistribution coefficient | 0.05 | 1.56 | Reference: chosen to expose the mechanism; under the time and length units it corresponds to 2.0 μm$^2$ s$^{-1}$, about 1/31 of the tortuosity-corrected 62.3 μm$^2$ s$^{-1}$ (Waters 2010). Mapped: set from 62.3 μm$^2$ s$^{-1}$ (Waters 2010). | controls spatial smoothing of monomer |
| $\lambda$ | monomer restoration/clearance coefficient | 0.30 | 126.7 | Reference: chosen equal to $r_T$, so that monomer evolves on the tau time scale; corresponds to 0.6 month$^{-1}$, about 1/420 of the measured value. Mapped: set from the interstitial Aβ half-life of 1.97 h, $9.8 \times 10^{-5}$ s$^{-1}$ (Cirrito et al. 2003). | restores monomer toward background |
| $k_E$ | fibril elongation coefficient | 4.0 | 15 | Reference: chosen so that tau-exposed tissue becomes elongation-controlled ($Da_{EC}$ = 15.6 at the source at detachment, 2.5–7.6 at the moving maximum; Sections 5.1, 5.5). Mapped: chosen so that $Da_{EC}$ in tau-exposed tissue is comparable to the 1.2–2.3 range derived from microdialysis data ((Hong et al. 2011); Section 5.5); in the mapped run $Da_{EC}$ is about 1.0 at the moving maximum and rises at the source from 3.3 at 4 months to 7.9 at 15 months. The arbitrary scale of $P$ is absorbed into $k_E$, so the measured elongation coefficient (Mosconi et al. 2026) enters only through $2k_EP$. | creates the elongational monomer sink |
| $k_\tau$ | heterotypic tau–Aβ nucleation coefficient | 0.50 | 0.50 | Chosen. Tau catalysis of Aβ42 primary nucleation, and its dependence on seed concentration and polymorph, is established *in vitro* (Mosconi et al. 2026); see Section 2.5. Its strength in tissue relative to $k_2$ has not been measured. | couples tau to Aβ fibril birth |
| $k_2$ | homotypic secondary-nucleation coefficient | 0.80 | 0.80 | Chosen. Absolute $R_2$ is not calibrated, which would require active-site density and per-site activity measured under matched conditions; $k_2$ does not appear in the hotspot condition, Eq. (18). | sets the scale of $R_2$ |
| $\mu_P$ | effective fibril-number loss coefficient | 0.02 | 0.02 | Chosen small; no tissue measurement. Under the time unit it corresponds to a fibril-number half-life of about 17 months. | limits accumulation of $P$ |

| Parameter | Meaning | Reference value | Mapped value | Reference or estimation method | Role in the calculation |
|---|---|---|---|---|---|
| $\mu_M$ | fibril-mass loss coefficient | 0 | 0 | Deposited mass is retained over the simulated interval, consistent with plaques that persist and grow over months *in vivo* (Hefendehl et al. 2011), (Burgold et al. 2014). | controls persistence of deposited mass |
| $\nu$ | mass assigned per nucleation event | 0.02 | 0.02 | Chosen small, so that fibril mass is added mainly by elongation; nuclei contain few monomers. A critical nucleus of two was assumed in the fits (Mosconi et al. 2026). | adds a small birth contribution to $M$ |
| $M_c$ | shielding crossover mass | 0.20 | 0.20 | Chosen. Eq. (10) is a tissue-scale closure whose parameters have not been identified experimentally (Section 2.3); its large-deposit limit, $A$ proportional to $M^{2/3}$, is consistent with surface-limited plaque growth (Hefendehl et al. 2011), (Burgold et al. 2014). | sets the transition in $A(M)$ |

### 2.5. Choosing parameter values: measured, inferred, or exploratory?

The model parameters fall into three evidentiary categories. First, some quantities are measured directly or constrained experimentally: an Aβ42 study (Hu et al. 2026) measured approximately one Brichos binding site per 143 ± 16 fibrillar Aβ42 monomers; independent microdialysis studies (Cirrito et al. 2003) and (Hong et al. 2011) measured an interstitial Aβ half-life near 2 h in plaque-free mice; and a transport study (Waters 2010) reported diffusible extracellular Aβ42 concentrations of roughly 160–425 pM together with an effective tissue diffusivity of $0.623 \times 10^{-6}$ cm$^2$ s$^{-1}$ after a tortuosity correction. Second, some quantities are inherited kinetic estimates: the reported elongation coefficient (Mosconi et al. 2026) is $3.5 \times 10^8$ M$^{-1}$ h$^{-1}$ = 0.0972 μM$^{-1}$ s$^{-1}$, and their seed-dependent primary-nucleation coefficients constrain the relative strength of PHF- and CTE-tau catalysis. Third, several reduced coefficients in Table 1 are exploratory because no tissue measurement fixes them independently. The reference and mapped columns make that distinction explicit.

### 2.6. Mapping the reduced model to physical time and length scales

Time-scale separation allows the monomer equation of the reduced model, Eq. (12), to be replaced by its quasi-steady form in the physically mapped calculations. In dimensional notation, free Aβ monomer relaxes at the local rate $\lambda_m + 2k_+P$, where $\lambda_m$ is monomer clearance and $2k_+P$ is the elongational sink per unit monomer. Using the microdialysis clearance rate together with the plaque-associated sink estimated in Section 5.5 gives a monomer relaxation time of roughly 0.9–1.3 h. Tau evolves much more slowly: the reported replication rates (Meisl et al. 2021) correspond to e-folding times of about 1.7 months in P301S mice and 7.1 years in human disease. Fibril number and mass also change on the slow time scale. Therefore, in the physically mapped calculations, the monomer field can be treated as the near-instantaneous response to the current tau and fibril distributions.

Setting $\partial m/\partial t = 0$ in Eq. (12) gives

$$0 = D_m \frac{\partial^2 m}{\partial x^2} + \lambda(1 - m) - 2k_E mP - \nu(R_\tau + R_2) \tag{16}$$

Eq. (16) introduces no new parameters; it uses the reduced quantities already defined after Eqs. (11)–(15). Its purpose here is to expose the spatial screening scale. Linearizing the monomer balance gives the full screening length $\ell_m = [D_m/(\lambda + 2k_E P)]^{1/2}$, the distance over which diffusion can resupply monomer against both background restoration/clearance and elongation. Here $\lambda + 2k_E P$ is the reduced-variable form of the dimensional relaxation rate $\lambda_m + 2k_+ P$ used above. The related elongation-only length $\ell_E = [D_m/(2k_E P)]^{1/2}$ isolates elongation versus diffusion. With the independently constrained transport and sink estimates used in Section 5.5, $\ell_m$ is approximately 0.44–0.54 mm.

The dimensional mapping is performed in two stages. First, $r_T$ and $D_T$ are matched to the measured time and length scales of tau propagation. Matching $r_T = 0.30$ to a tau replication rate of 0.6 month$^{-1}$ gives a model time unit of 0.50 months; matching $D_T = 0.0025$ to an effective spreading coefficient of order $10^{-13}$ m$^2$ s$^{-1}$ gives $L$ = 7.20 mm as the model length unit. Under this tau-based mapping, the $T = 0.1$ contour moves at about 0.79 mm per month. Second, the physically interpreted monomer calculations use the measured Aβ diffusivity and clearance scale, together with the plaque-associated sink inferred in Section 5.5. This second step is necessary because the exploratory reference values of $D_m$, $\lambda$, and $k_E$ were deliberately chosen to make the mechanism visible rather than to reproduce tissue kinetics.

The distinction between the two parameter sets is therefore central to interpreting the Results. Sections 5.1–5.4 use the reference system to establish the mechanism and verify the mathematics. Sections 5.5–5.8 ask whether the controlling ratios and length scales are compatible with measurements. The mapped calculation provides a mechanistic dimensionalization rather than a patient-specific fit. The two most robust physical quantities are the kinetic Damköhler number, which is of order unity or larger in deposit-rich tissue, and the monomer screening length, which is of order half a millimeter.

## 3. Analytical result: what determines the hotspot position?

### 3.1. The competition is between catalytic sites and usable monomer

The reduced model makes the mechanism transparent. Tau increases local fibril birth, which raises $P$, $M$, and the accessible catalytic sites $A(M)$. But the accompanying increase in $P$ creates more elongating ends and lowers $m$. Homotypic secondary nucleation therefore depends on the product of two fields that change in opposite directions: $R_2 = k_2 A(M) m^2$. Once monomer becomes limiting near the tau-exposed source, moving away from that source can increase $R_2$ even while $A(M)$ is decreasing.

### 3.2. An exact gradient balance locates any interior hotspot

Let $x_h$(t) denote the position of an interior maximum of $R_2$ at time $t$. Because $R_2 = k_2 A(M) m^2$ and $k_2$ is spatially uniform, differentiating ln $R_2$ with respect to $x$ gives

$$\frac{1}{R_2}\frac{\partial R_2}{\partial x} = \frac{\partial \ln A}{\partial x} + 2\frac{\partial \ln m}{\partial x} \tag{17}$$

At an interior maximum, $\partial R_2/\partial x = 0$. Therefore $x_h(t)$ must satisfy

$$\left[\frac{\partial \ln A}{\partial x}\right]_{x_h} = -2\left[\frac{\partial \ln m}{\partial x}\right]_{x_h} \tag{18}$$

Eq. (18) is the central spatial result. Moving away from an old tau-exposed region, accessible catalytic sites typically decrease, so $\partial x$ ln $A < 0$, while free monomer recovers, so $\partial x$ ln $m > 0$. The hotspot lies where those opposing logarithmic gradients exactly balance. Because monomer enters as $m^2$, its logarithmic gradient is weighted by a factor of two. The condition is local and does not depend on the particular equation chosen for tau spreading.

### 3.3. Two Damköhler numbers separate the two transport competitions

The first dimensionless ratio asks whether elongation can deplete monomer faster than background restoration can replenish it. In the reduced model we define

$$\mathrm{Da}_{\mathrm{EC}}(x,t) = \frac{2k_E P(x,t)}{\lambda} \tag{19}$$

In Eq. (19), $\mathrm{Da}_{\mathrm{EC}}(x,t)$ is the local elongation-to-clearance Damköhler number; $k_E$ is the reduced elongation coefficient, $P(x,t)$ is the local fibril concentration, and $\lambda$ is the monomer restoration/clearance coefficient. $\mathrm{Da}_{\mathrm{EC}} \ll 1$ means restoration dominates, whereas $\mathrm{Da}_{\mathrm{EC}} \gg 1$ means elongation controls the local monomer pool. A second ratio asks whether diffusion can smooth that depletion over a chosen tissue length $L$:

$$\mathrm{Da}_D(x,t) = \frac{2k_E P(x,t) L^2}{D_m} \tag{20}$$

In Eq. (20), $\mathrm{Da}_D(x,t)$ is the elongation-to-diffusion Damköhler number, $L$ is the tissue length over which redistribution is being assessed, and $D_m$ is the monomer redistribution coefficient. $\mathrm{Da}_D \ll 1$ means diffusion smooths monomer faster than elongation can sustain a gradient over $L$; $\mathrm{Da}_D \gg 1$ means the depletion can persist. An off-source hotspot therefore requires both a sufficiently large $\mathrm{Da}_{\mathrm{EC}}$ in the wake and a sufficiently large $\mathrm{Da}_D$ on the spatial scale of interest.

The same spatial competition can be expressed as lengths. Rearranging Eq. (20) gives the elongation-only length $\ell_E = [D_m/(2k_E P)]^{1/2}$ and $\mathrm{Da}_D = (L/\ell_E)^2$. Including background restoration gives the shorter full screening length $\ell_m = [D_m/(\lambda + 2k_E P)]^{1/2} = \ell_E[\mathrm{Da}_{\mathrm{EC}}/(1 + \mathrm{Da}_{\mathrm{EC}})]^{1/2}$. $\ell_E$ isolates elongation versus diffusion; $\ell_m$ is the physical distance over which a monomer perturbation persists. In the elongation-controlled limit, the two lengths approach one another. These are the same two lengths introduced after Eq. (16).

### 3.4. Hotspot motion and tau-front motion are different quantities

The Fisher–Kolmogorov equation supplies a reference propagation speed, $c_T = 2(D_T r_T)^{1/2}$, but the secondary-nucleation maximum is not a traveling-wave solution of Eq. (1). Its position depends on the accumulated histories of $P$, $M$, and $m$. Throughout the Results, "tau front" means the $T = 0.1$ contour unless stated otherwise. Because the Fisher–Kolmogorov field has a diffuse positive leading tail, a hotspot can lie ahead of that contour while still occupying tissue with nonzero $T$. A numerical lead over the contour must therefore not be interpreted as Aβ activation propagating through tau-free tissue.

### 3.5. Which model quantities have experimental counterparts?

The theory is organized around $R_2$, the instantaneous homotypic secondary-nucleation rate, but no current tissue assay measures $R_2$ directly. Amyloid PET primarily reports fibrillar burden $M$; longitudinal PET can estimate changes in burden analogous to $\partial M/\partial t$. The molecular fibril-birth rate $R_\tau + R_2$ is closer to nucleation activity but is not identical to the rate of appearance of macroscopic plaques, because many fibrils can contribute to one deposit. Free monomer $m$ can be sampled by microdialysis, although only as a tissue average. The Results therefore track these quantities separately rather than treating any one of them as a direct proxy for another.

## 4. Numerical solution and verification

The dimensionless reference system, Eqs. (11)–(15), was solved on $0 \leq x \leq 1$. No-flux boundary conditions were imposed on $T$ and $m$. Initially, tau was localized near $x = 0$ as $T(x,0) = 0.02 \exp[-(x/0.04)^2]$, while $m(x,0) = 1$ and $P(x,0) = M(x,0) = 0$. Spatial derivatives were approximated with second-order centered finite differences. The resulting stiff method-of-lines system was integrated with a variable-order backward differentiation formula. Unless otherwise stated, the calculations use the reference parameter values in Table 1.

Numerical convergence was checked before interpreting hotspot motion. Calculations were repeated on 81-, 161-, and 321-point grids. Hotspot position was recorded at the grid point of maximum $R_2$ and was also estimated by parabolic sub-grid interpolation. Between 161 and 321 points, the largest change in hotspot position over the reported times was 0.0031 domain lengths at grid resolution and 0.0006 after interpolation; the peak $R_2$ changed by less than 0.1%. The 161-point grid was therefore retained for the mechanistic simulations.

The code can solve either the full time-dependent reduced system or the quasi-steady monomer form of Eq. (16). The elliptic monomer equation is solved by damped Newton iteration on a sparse tridiagonal Jacobian and converges in at most three iterations per evaluation. The reference calculation keeps $\partial m/\partial t$ because its intentionally balanced time scales do not justify a quasi-steady approximation. Under the mapped physical ratio $\lambda/r_T \approx 422$, the maximum difference between the quasi-steady and time-dependent monomer fields falls to $7.3 \times 10^{-4}$, confirming that Aβ monomer is effectively slaved to the slower tau/fibril dynamics in the physically interpreted regime.

**5. Results: from a tau source to a monomer-depleted wake**

**5.1. Tau first amplifies nucleation at the source; an outer maximum later takes over**

The reference calculation begins in the intuitive regime: both $T$ and $R_2$ are maximal at the source $x = 0$. The source value of $R_2$ reaches its global maximum near $t = 3.1$ while monomer is still relatively abundant. Continued tau-driven fibril production then increases $P$, so elongational demand grows. As $m$ falls, $R_2$ at the source declines even though fibril mass and accessible catalytic sites continue to increase. By $t \approx 8.3$ (about 4.2 months), the source-region $Da_{EC}$ has reached about 15.6 and a second, off-source maximum becomes larger than the source maximum. The oldest tau-exposed region has accumulated the most catalytic sites but no longer has enough substrate to use them most effectively.

The transition is an exchange of dominance between two maxima, not initially a single peak sliding away from the source. Before $t \approx 8.3$, a source maximum and an outer maximum coexist. The outer maximum then overtakes the source maximum. This distinction matters when interpreting finite domains: on the short reference domain, confinement eventually affects the outer peak, whereas on a longer mapped domain the outer maximum remains interior and behaves as a translating band.

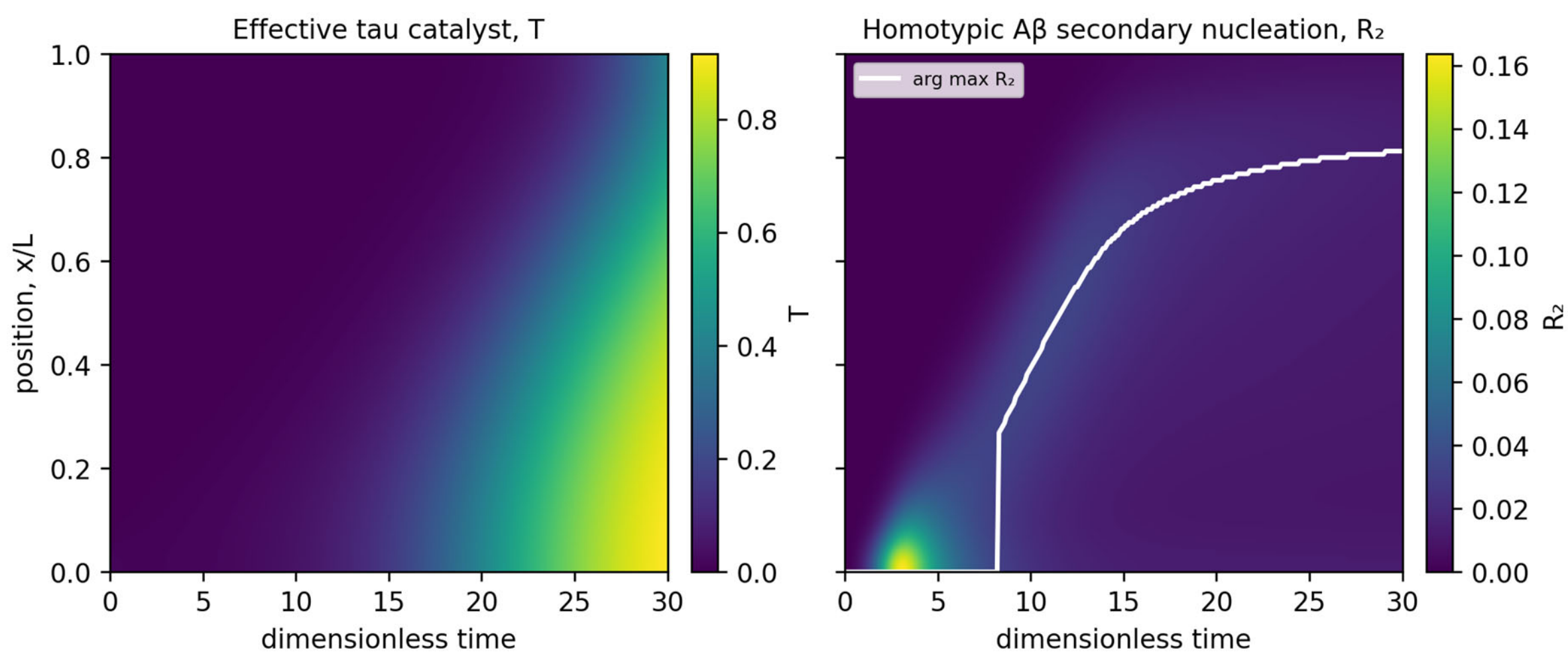

**Fig. 2. Space–time evolution of the effective tau catalyst $T$ (top) and the homotypic Aβ secondary-nucleation rate $R_2$ (bottom) in the reference dimensionless calculation. The solid white curve is the location $x_h(t)$ = arg max$_x$ $R_2$ and the dashed white curve is the outermost $T = 0.1$ contour, used here as the operational tau front. $R_2$ is bimodal: a maximum at the source and an outer maximum coexist, and the abrupt change near $t = 8$ is the outer maximum overtaking the inner one rather than a single peak migrating continuously. In this confined domain the $T = 0.1$ contour later overtakes the nucleation maximum; Section 5.6 shows how that ordering changes when boundary confinement is removed.**

### 5.2. Why the outer maximum wins: fewer sites, but much more monomer

After the outer maximum becomes dominant, its location moves progressively away from the source: $x_h/L$ = 0.394 at $t$ = 10, 0.525 at $t$ = 12, 0.663 at $t$ = 15, 0.756 at $t$ = 20, 0.794 at $t$ = 25, and 0.813 at $t$ = 30. A least-squares fit over $t$ = 10–30 gives 0.0171 domain lengths per unit time in the confined reference calculation. That speed is not the central result. The important point is that $x_h$ follows the accumulated balance between $A(M)$ and $m$ rather than the instantaneous tau field $T$.

The mechanism is especially clear at $t$ = 12. At the hotspot, $x_h/L$ = 0.525, accessible catalytic capacity is only 0.283, about 14% of its source value 1.976. Free monomer, however, is 0.386, almost four times the source value 0.105. Because $R_2 \propto A(M)m^2$, the gain in monomer more than compensates for the loss of accessible catalytic sites, and the off-source rate is nearly twice the source rate. This comparison captures the central mechanism in one sentence: the region with the most accessible catalytic sites can have a lower nucleation rate than a region with fewer sites because it has depleted the substrate required to use those sites.

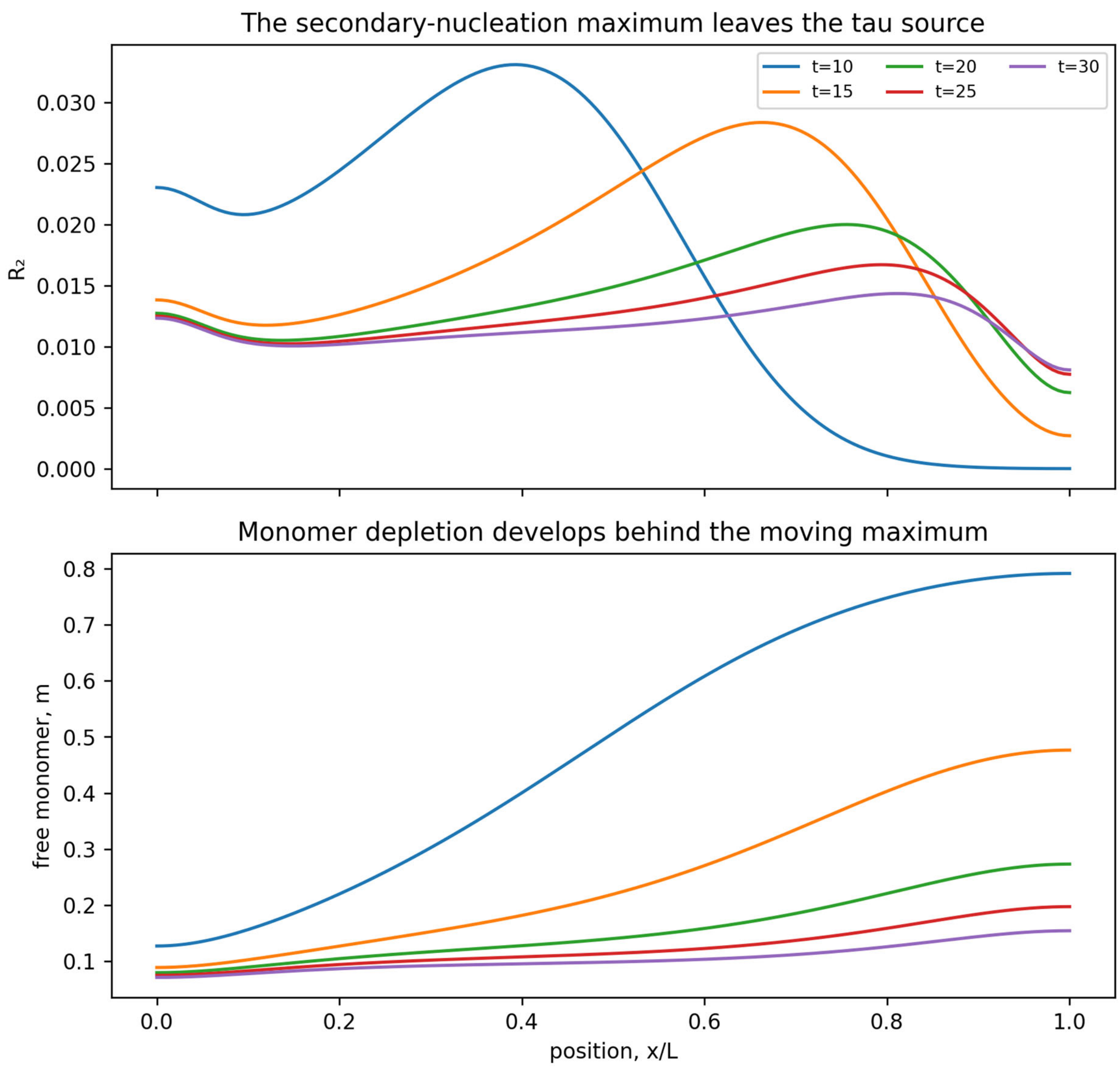

**Fig. 3. Spatial profiles of homotypic secondary nucleation $R_2$ (top) and free Aβ monomer $m$ (bottom) at selected times. The $R_2$ maximum moves away from the tau source at $x = 0$. Behind it, larger fibril and active-site burdens are offset by stronger monomer depletion.**

### 5.3. The numerical maximum obeys the analytical gradient balance

Eq. (18) makes a quantitative prediction: at an interior maximum, the negative logarithmic gradient of accessible catalytic sites must equal twice the positive logarithmic gradient of monomer. The numerical solution satisfies this relation closely. From $t = 10$ to 30 the two sides agree on average within about 1% (at most 2.4%) when evaluated at the grid maximum and within 0.11% after parabolic sub-grid localization. The remaining discrepancy is therefore a localization error, not a failure of the mechanism. The moving maximum is the point where the loss of sites and recovery of monomer balance each other exactly as predicted.

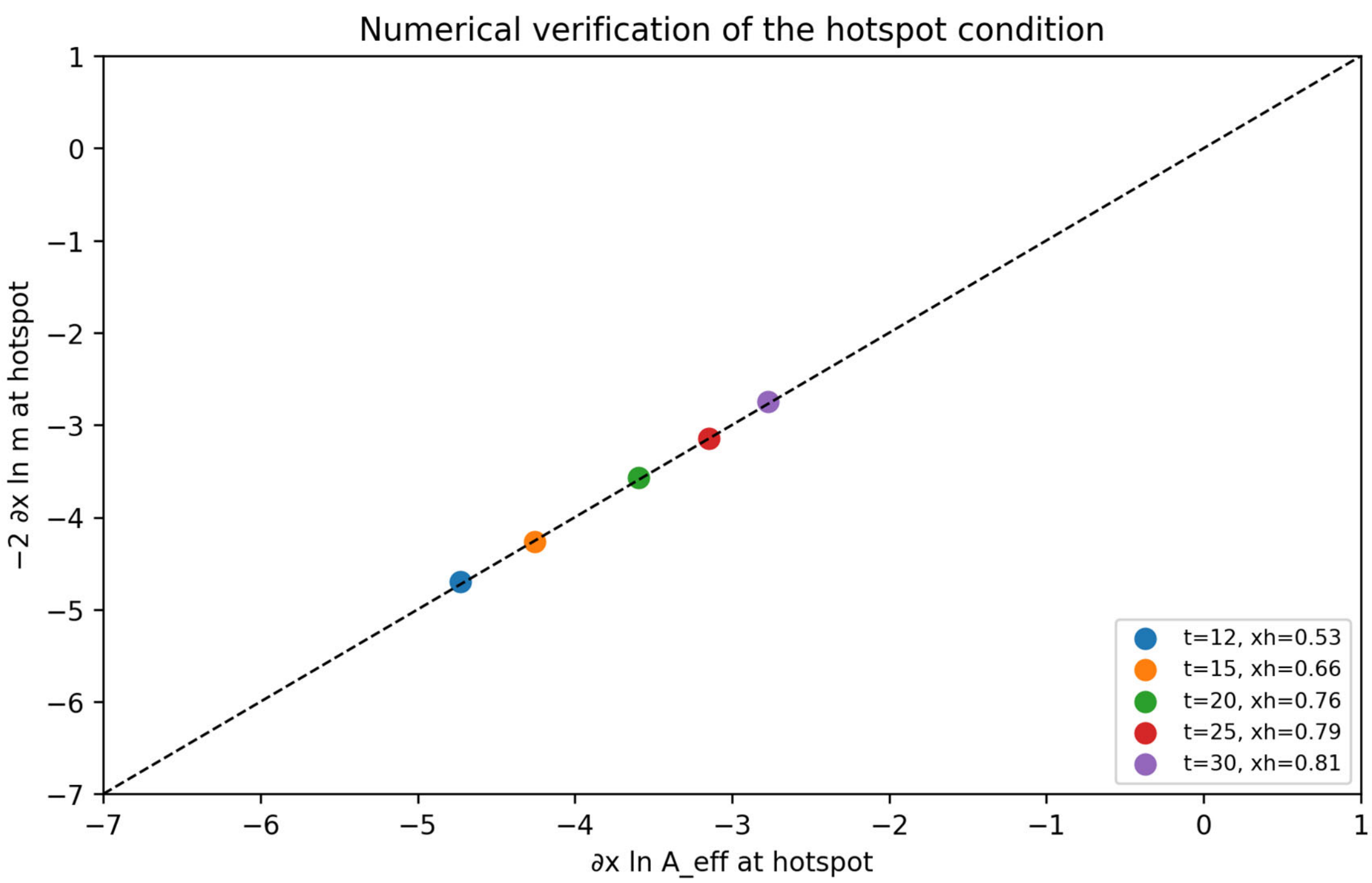

**Fig. 4. Verification of Eq. (18). Each symbol compares the two logarithmic gradients evaluated at the numerical hotspot for $t$ = 12, 15, 20, 25, and 30. The dashed line is equality. The close collapse demonstrates that the hotspot is governed by the balance between accessible catalytic sites and the squared monomer factor.**

### 5.4. Diffusion can erase the hotspot before it forms

The hotspot mechanism requires a spatial monomer gradient. We therefore vary $D_m$ while keeping all other reduced parameters fixed. Slower redistribution allows depletion to persist and causes earlier, larger displacement; faster redistribution delays the transition. For $D_m = 0.01$, 0.02, 0.05, and 0.10, the maximum leaves the source at $t = 4.4$, 5.5, 8.3, and 12.8, respectively. At $t = 20$ the corresponding position shifts inward from $x/L = 0.900$ to 0.625 as $D_m$ increases. At $D_m = 0.20$, the off-source maximum never becomes dominant during the simulated interval. This sensitivity test isolates causality within the model: without a persistent monomer gradient, there is no displaced hotspot.

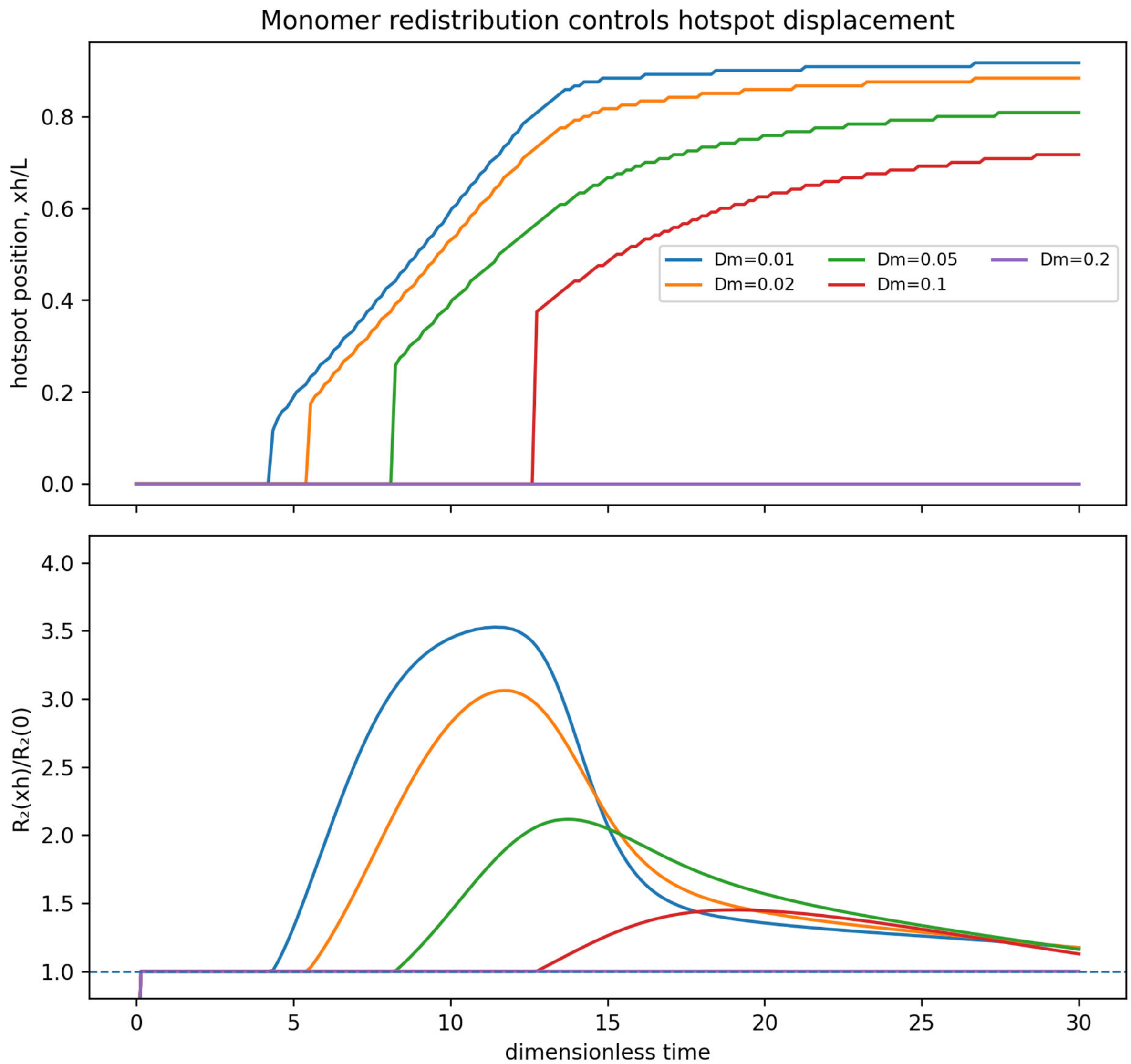

**Fig. 5. Sensitivity to the effective monomer-redistribution coefficient $D_m$. Top: hotspot position. Bottom: hotspot rate normalized by the rate at the tau source. Slow to intermediate redistribution permits the maximum to detach and move; sufficiently rapid redistribution ($D_m$ = 0.20) keeps the maximum at the source. As in Section 5.1, the transition at every $D_m$ is an outer maximum overtaking the source maximum rather than a single peak migrating continuously, which is why the plotted position jumps at detachment.**

### 5.5. The physical prediction is mesoscale: a screening length of roughly half a millimeter

The two Damköhler numbers now acquire a direct physical interpretation. The values quoted in this paragraph are still those of the reference calculation of Sections 5.1–5.4; the independent measurements follow below. Along the reference hotspot trajectory, $\mathrm{Da}_{\mathrm{EC}}$ increases from 2.50 at $t$ = 10 to 7.55 at $t$ = 30, so elongation is already competitive with or stronger than monomer restoration. Over the full reference domain, $\mathrm{Da}_D$ increases from 15.0 to 45.3, so diffusion is not fast enough to erase the depletion field. The hotspot therefore sits between two kinetic histories: a strongly depleted wake behind it and a region ahead that contains more monomer but fewer catalytic sites. It is not simply the contour $\mathrm{Da_{EC}} = 1$.

Independent microdialysis data place plaque-rich tissue in a comparable elongation-to-clearance regime. A microdialysis study (Hong et al. 2011) reported that recovery of injected $^{125}$I-labeled Aβ1–40 from interstitial fluid in old, plaque-rich J20 mice was about 45% of that in young, plaque-free mice. Because no corresponding first-order rate constant was reported, we estimate the additional plaque-associated sink in three ways. Interpreting the reduction to 45% recovery over the 1-h collection interval as first-order loss gives about 0.8 $\mathrm{h^{-1}}$ ($2k_E P \approx 2.2 \times 10^{-4}\ \mathrm{s^{-1}}$); treating recovery as inversely proportional to total removal, using the 1.97 h plaque-free half-life, gives about 0.43 $\mathrm{h^{-1}}$ ($1.2 \times 10^{-4}\ \mathrm{s^{-1}}$); and the steady-state decline in interstitial Aβ gives approximately $1.4 \times 10^{-4}\ \mathrm{s^{-1}}$. Relative to the plaque-free clearance rate of $9.8 \times 10^{-5}\ \mathrm{s^{-1}}$ (Cirrito et al. 2003), these estimates correspond to $\mathrm{Da_{EC}} \approx 1.2$–2.3. Thus elongation can be a substantial monomer sink in deposit-rich tissue.

The corresponding spatial scale is larger than the spacing between individual plaques. With $D_m = 62.3\ \mu\mathrm{m^2\ s^{-1}}$, the elongation-only length $\ell_E$ is approximately 530–720 μm and the full screening length $\ell_m$ is approximately 440–535 μm. Plaque imaging (Burgold et al. 2014) measured a mean nearest-neighbor plaque spacing of only 65.8 μm in Tg2576 mice. Taking that spacing as $L$ gives $L/\ell_m \approx 0.12$–0.15 on that scale. Monomer should therefore be nearly uniform between neighboring plaques. The predicted displacement is not a micrometer-scale ring around a single plaque; it is a mesoscale tissue phenomenon requiring distances of order half a millimeter.

This distinction also tells us which experimental observations are relevant. Plaque-local oligomer halos (Koffie et al. 2009) and new-plaque clustering within tens of micrometers (McCarter et al. 2013) occur on scales far below $\ell_m$ and can arise from local seeding, transport, or microglial effects. The present mechanism becomes important only on a scale large enough for a monomer deficit to persist against diffusion. Section 5.4 and the dimensional estimate therefore express the same requirement in two ways: rapid exchange on the scale of interest eliminates the displacement.

### 5.6. Mapping the model to physical units: what is robust and what is not?

The first dimensional mapping uses the tau field to place the reference geometry on a physical axis. Under that mapping, the unit domain becomes a 7.20-mm segment, on which the outer maximum leaves the source at about 4.2 months while the $T = 0.1$ contour advances at 0.79 mm per month. These positions are useful for identifying when the finite boundary begins to influence the solution, but they are not a quantitative mouse prediction: the reference monomer coefficients are exploratory rather than physiological.

Independent measurements are therefore used as consistency checks on the mechanism, not as evidence of a fitted model. The calculated source-region monomer depletion is of the same order as the decline in interstitial

Aβ42 observed as plaques accumulate (Hong et al. 2011). The model also allows nucleation activity to peak and decline while fibril mass continues to rise, qualitatively echoing the age-dependent decline in new-plaque formation (Hefendehl et al. 2011) and (Burgold et al. 2014). A transgenic study (Pooler et al. 2015) provides a spatially staged tau–amyloid observation, but because that experiment does not measure secondary nucleation, it supports only the plausibility of localized coupling.

A negative comparison is equally informative. The maximum of $\partial M/\partial t$ remains at the tau source in both the reference and mapped calculations. Amyloid PET, which primarily reports fibrillar burden and its longitudinal change, therefore does not directly report the model's $R_2$ maximum. The regional reorganization of amyloid accumulation described previously (Guo et al. 2018) is not reproduced by this homogeneous model and should not be presented as validation. By contrast, the molecular fibril-birth field $R_\tau + R_2$ closely follows $R_2$, so new-deposit formation is a possible qualitative—but not one-to-one—experimental proxy.

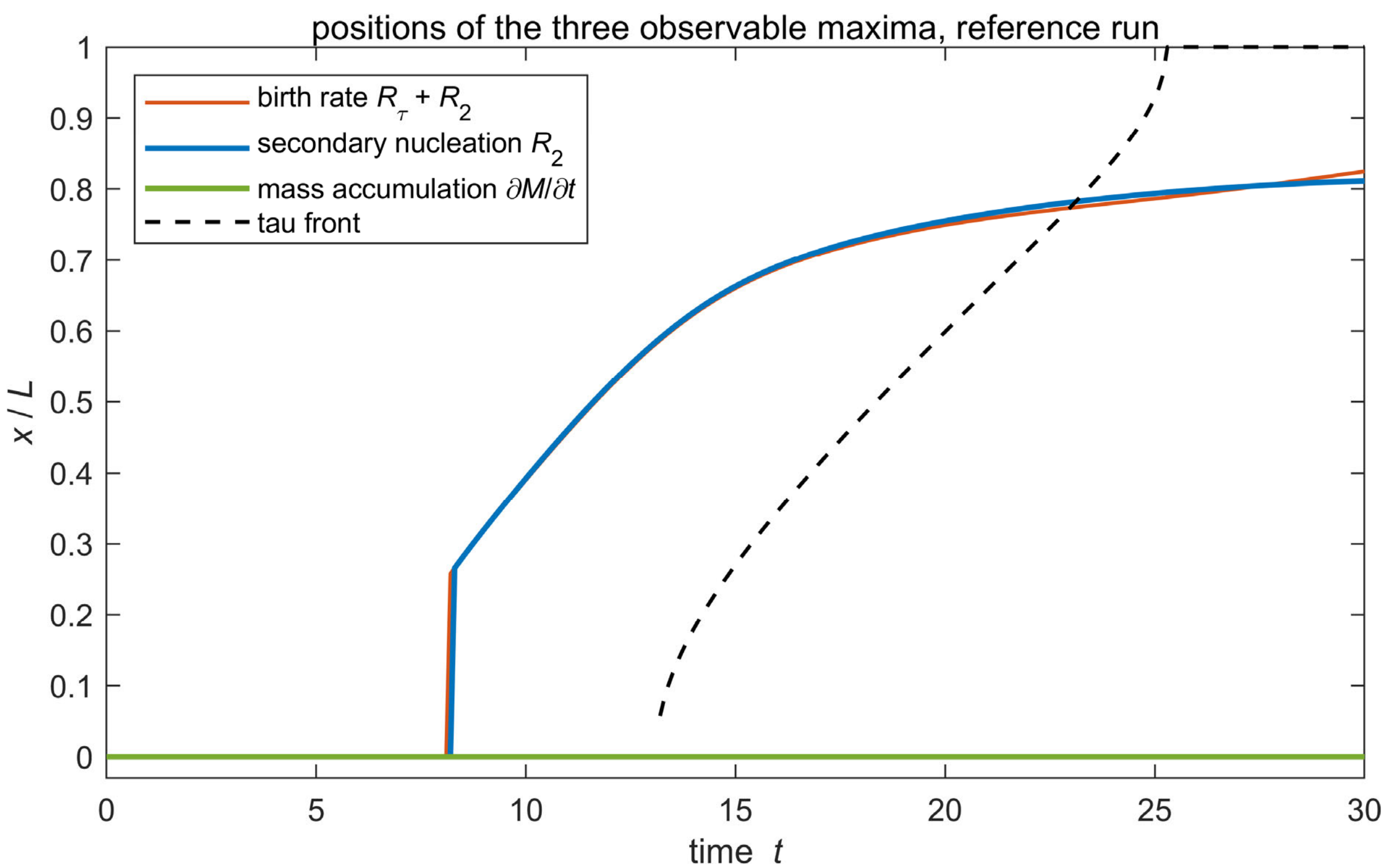

**Fig. 6. Positions of the three model maxima and the tau front defined in Section 3.4 for the reference calculation: the rate of fibril birth $R_\tau + R_2$, the homotypic secondary-nucleation rate $R_2$, the rate of fibril-mass accumulation $\partial M/\partial t$, and the tau front. The birth rate and the secondary-nucleation rate are indistinguishable at this scale, which motivates considering the appearance of new deposits as a qualitative proxy for a quantity that cannot itself be measured directly. The maximum of $\partial M/\partial t$ remains at the source at every time.**

Boundary confinement mainly affects the apparent speed of the outer maximum. Repeating the mapped calculation on longer segments at a fixed physical resolution gives a converged outer-maximum speed of 1.650 mm per month on 30-, 40-, and 50-mm segments, compared with 1.562 mm per month on a 20-mm segment where the boundary is still felt. The operational $T = 0.1$ contour moves at 0.79 mm per month, so the unconfined outer maximum leads that particular contour over the simulated interval. This numerical ratio is contour dependent because the Fisher–Kolmogorov field has a diffuse leading tail; the maximum remains in tissue with nonzero *T*. The robust result is therefore spatial separation, not a universal speed ratio.

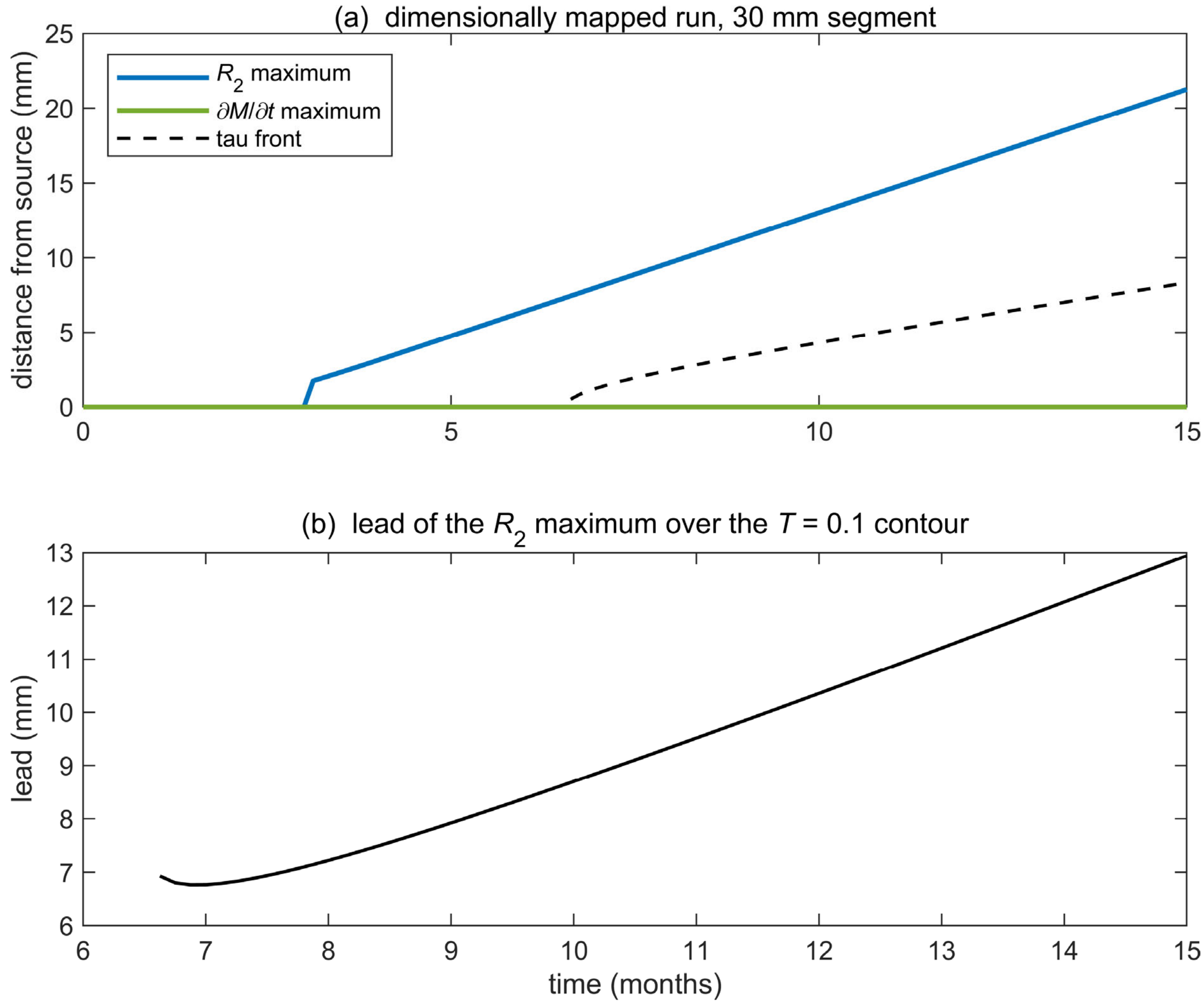

**Fig. 7. Dimensionally mapped calculation on a 30-mm segment. (a) Position of the secondary-nucleation maximum, the mass-accumulation maximum, and the operational tau front defined by *T* = 0.1. (b) Separation between the nucleation maximum and the *T* = 0.1 contour, which grows to about 13 mm by 15 months. The spatial resolution is held at 0.05 mm so that the approximately 0.66 mm tau-front width remains resolved; the outer-maximum speed, fitted over 5–15 months, agrees to three decimal places with the values obtained on the 40- and 50-mm segments. The lead is contour dependent because the Fisher–Kolmogorov field has a diffuse leading tail.**

**Table 2. Model quantities compared with independent experimental estimates or qualitative observations. Model entries are from the dimensionally mapped calculation on a 30-mm segment (Section 5.6) unless stated otherwise.**

| Quantity | Model | Independent measurement |
|---|---|---|
| Operational *T* = 0.1 tau-contour speed | 0.79 mm per month | about 0.6–1.2 mm per month, estimated here from pathology maps after injection of AD-brain tau into mouse hippocampus and overlying cortex (He et al. 2018) |
| Monomer depletion at the tau source | 89% below background at 15 months | 80% fall in interstitial Aβ42 between 3 and 24 months (Hong et al. 2011) |
| Kinetic Damköhler number in deposit-rich tissue | about 1.0 at the moving maximum; 3.3 to 7.9 at the source from 4 to 15 months | 1.2 to 2.3, derived here from tracer recovery and from steady-state interstitial Aβ (Hong et al. 2011) |
| Monomer screening length | approximately 440–535 μm, computed from the measured inputs in the next column rather than predicted independently | approximately 440–535 μm from measured diffusivity, clearance, and plaque-associated sink (Waters 2010), (Cirrito et al. 2003), (Hong et al. 2011) |
| Order of secondary nucleation in monomer | 2 in Eqs. (2)–(4); supported by fits near 2 | close to 2 when fitted as a free parameter (Mosconi et al. 2026) |
| Scaling of accessible catalytic sites | $A$ proportional to $M^{2/3}$ for large deposits | similar radial growth of pre-existing and newly formed plaques in APPPS1 mice, 0.286 versus 0.306 μm per week (Hefendehl et al. 2011) |
| Nucleation declines while fibril mass rises | source-region $R_2$ peaks at 1.9 months | new-plaque formation peaks at 4 to 5 months of age in APPPS1 mice and then declines (Hefendehl et al. 2011); 3 new versus 786 pre-existing plaques in aged Tg2576 mice, compared with 50 new versus 101 pre-existing in young mice (Burgold et al. 2014) |
| Enhanced deposition follows the arrival of tau | off-source maximum emerges after tau exposure; dimensional position is domain- and contour-dependent | plaque area raised in entorhinal cortex at 10 months and in somatosensory cortex only by 16 months (Pooler et al. 2015) |

### 5.7. Oligomer transport tests whether the spatial signal can remain local

The reduced model sends each nucleation event directly into the local fibril population. To test whether that simplification matters, we reintroduce $S(x,t)$, the soluble oligomer-equivalent field defined in Section 2.2. A capillary study (Peter et al. 2026) showed that Aβ42 aggregation can be transmitted by oligomers diffusing ahead of a fibril-growth front. If newly generated oligomers travel far before converting to fibrils, they can smear out the spatial signal required by Eq. (18). We therefore solve the extended balances

$$\frac{\partial S}{\partial t} = D_S \frac{\partial^2 S}{\partial x^2} + R_\tau + R_2 - k_{\text{conv}} S \tag{21}$$

$$\frac{\partial P}{\partial t} = k_{\text{conv}} S - \mu_P P, \qquad \frac{\partial M}{\partial t} = 2k_E m P + \nu k_{\text{conv}} S - \mu_M M \tag{22}$$

In Eqs. (21)–(22), $D_S$ is the oligomer diffusion coefficient and $k_{\text{conv}}$ is the first-order conversion rate from soluble oligomer equivalents $S$ to fibrils. The coefficients $\mu_P$ and $\mu_M$ are the fibril-number and fibril-mass loss terms already defined in the reduced model, while $\nu$ converts each oligomer-to-fibril conversion event into fibrillar mass. These equations introduce one new transport length, $\ell_S = (D_S/k_{\text{conv}})^{1/2}$, the characteristic distance an oligomer diffuses before conversion. We measure the surviving spatial structure by the hotspot contrast, defined as the value of $R_2$ at the moving maximum divided by its value at the tau source. The results are threshold-like: below about 0.15 mm the hotspot contrast is essentially unchanged; between 0.16 and 0.30 mm it collapses; and above 0.5 mm $R_2$ is nearly uniform across the long domain. A conversion delay without diffusion changes the hotspot position modestly but leaves the contrast nearly unchanged, showing that transport—not delay—is responsible for the loss of spatial structure.

Current observations do not yet determine the effective tissue value of $\ell_S$. The reported oligomer diffusivity (Peter et al. 2026), 0.664 mm$^2$ h$^{-1}$, divided by the square of an assumed tissue tortuosity of 1.7 and combined with their adopted conversion rate $3.3 \times 10^{-2}$ h$^{-1}$, gives $\ell_S \approx 2.64$ mm—large enough to erase the hotspot in this model. In fixed APP/PS1 mouse cortex, however, histological imaging (Koffie et al. 2009) showed an oligomer-rich halo extending only about 6.5 μm beyond dense plaque cores, although smaller deposits were also present farther away. That halo is not a direct measurement of $\ell_S$ because binding, production, and clearance also shape it. The relevant experimental target is therefore the effective transport length of nucleation products in tissue.

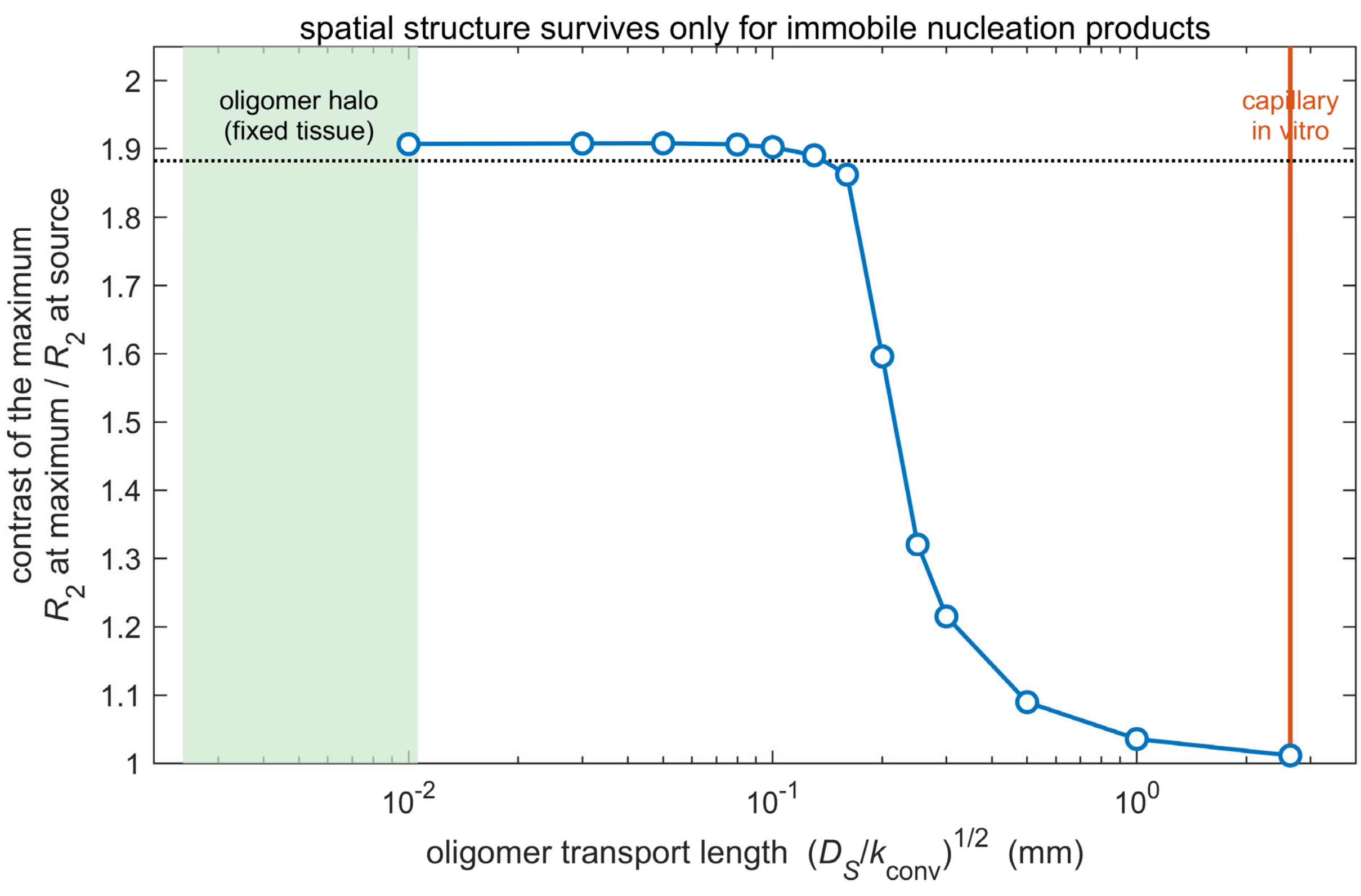

**Fig. 8. Survival of the spatial structure against oligomer transport. The contrast of the maximum, $R_2$ at the maximum divided by $R_2$ at the source, is plotted against the oligomer transport length $(D_S/k_{conv})^{1/2}$, with all other parameters at their dimensionally mapped values on a 30-mm segment. The dotted line is the base-model contrast. The shaded band marks the much smaller plaque-local scale over which oligomeric Aβ is concentrated around plaque cores in fixed mouse cortex (Koffie et al. 2009); it is shown only as a qualitative localization reference, not as a direct estimate of $D_S/k_{conv}$. The vertical line is the value**

**obtained from the capillary measurements (Peter et al. 2026) after correction for tissue tortuosity. The modeled hotspot contrast collapses between transport lengths of about 0.16 and 0.30 mm.**

### 5.8. Two refinements act in opposite directions: reversible exchange weakens the hotspot, growth history strengthens it

The base model assumes that deposited Aβ acts as a persistent monomer sink and that catalytic-site density per unit fibril mass is fixed. We test both assumptions. First, we allow deposited Aβ to return to the soluble pool by replacing irreversible elongation with the net flux $2k_E mP - k_{\mathrm{off}}M$, where $k_{\mathrm{off}}$ is a first-order release coefficient. The corresponding local equilibrium monomer concentration is $k_{\mathrm{off}}M/(2k_E P)$. Because release replenishes the depleted wake, increasing $k_{\mathrm{off}}$ should reduce the hotspot contrast.

That is what the calculations show. The hotspot contrast falls from 1.88 in the irreversible model to 1.29 when the source release flux is half the production flux and to 1.03 when release and production are equal. Surface-limited exchange changes these values only modestly. Thus sufficiently strong sustained release can eliminate the displacement. The approximately twofold increase in apparent interstitial Aβ half-life observed in plaque-bearing PDAPP mice after Aβ production was blocked (Cirrito et al. 2003) does not by itself establish such sustained release over months; a small, rapidly exchanging surface pool could buffer monomer over hours while deposits remain a net sink over longer times.

The second refinement acts in the opposite direction. An Aβ40 study (Hu et al. 2026) found that slowly annealed Aβ40 fibrils formed near equilibrium had a catalytic-defect density 8- to 12-fold lower than that of more rapidly grown Aβ40 fibrils. To represent an analogous effect, we denote by $\sigma_{\mathrm{lo}}$ the site density of fibrils formed in the depleted wake, normalized to the reference site density. Using the Aβ40 result only as an illustrative sensitivity scale, lowering $\sigma_{\mathrm{lo}}$ increases the hotspot contrast from 1.88 to 2.41. The interpretation is straightforward: monomer depletion can suppress secondary nucleation twice—immediately by lowering $m$ and persistently by generating fibrils with fewer active sites per unit mass.

When reversible exchange and growth-history dependence are combined, their effects partly cancel. With a release flux equal to half the production flux and the illustrative Aβ40-based site-density dependence, the contrast is 1.61. Across the combinations tested, hotspot position changes by less than 1% even when hotspot amplitude changes substantially. The hotspot position is therefore more robust than its contrast. The two quantities that most need experimental constraint are the fraction of deposited Aβ that remains exchangeable over months and the dependence of catalytic-site density on the growth history of tau-seeded Aβ42 fibrils.

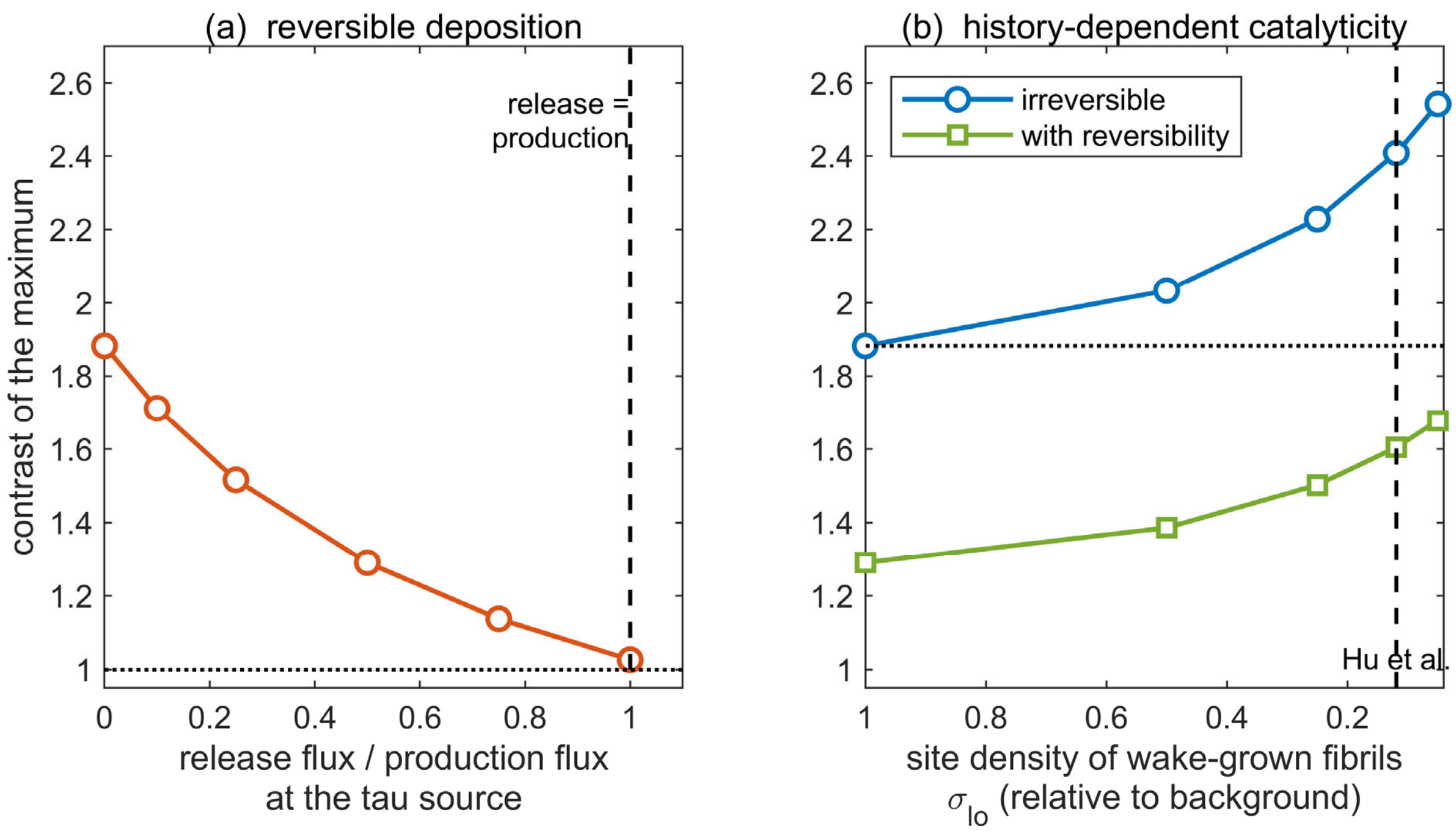

**Fig. 9. The two kinetic refinements act in opposite directions. (a) Contrast of the maximum against the ratio of the release flux to the production flux at the tau source; the dashed line marks release equal to production. (b) Contrast against the site density of fibrils grown in the depleted wake, $\sigma_{lo}$, without reversibility (circles) and with the release flux at half the production flux (squares); the dashed line marks the value implied by the 8.4-fold reduction in Brichos binding sites, at the low end of the 8- to 12-fold range reported for annealed Aβ40 fibrils (Hu et al. 2026), and the dotted line marks the base model. Decreasing $\sigma_{lo}$ to the left corresponds to a stronger history dependence.**

## 6. Discussion: what the moving hotspot means—and what it does not

The model answers the spatial paradox posed in the Introduction. If bioavailable tau locally catalyzes Aβ nucleation, it initially increases fibril number and accessible catalytic sites. Continued exposure, however, produces enough elongating fibril ends to deplete free monomer. The long-exposed tau region can therefore become a depleted wake, while a region farther along the direction of spread contains fewer catalytic sites but substantially more substrate. Because $R_2 = k_2 A(M) m^2$, the intermediate region can have the largest instantaneous secondary-nucleation rate. The predicted hotspot is thus a maximum of kinetic activity, not a simple copy of the tau or plaque-burden field.

This mechanism is the spatial counterpart of the companion lumped model (Kuznetsov 2026a). The earlier study showed that tau-induced fibril shortening is not sufficient to imply greater catalytic activity and that the homotypic response changes when monomer control shifts from clearance to elongation. Adding space allows those regimes to coexist: long-exposed tissue can be elongation dominated while newly reached tissue remains comparatively monomer-replete. The moving hotspot is the spatial expression of that Damköhler transition.

The most general result is Eq. (18), not the Fisher–Kolmogorov trajectory. Any process that creates a spatially varying accessible-site field and a spatially varying monomer field obeys the same local condition for an interior maximum of $R_2$. Network transport, anisotropic diffusion, axonal propagation, or explicit extracellular tau release could replace Eq. (1) without changing that gradient balance. Likewise, a different plaque-accessibility closure would change $A(M)$ but not the structure of the criterion.

This also distinguishes the present work from reaction–diffusion models that predict where total pathological burden accumulates. Fisher–Kolmogorov, heterodimer, network, and Gray–Scott models can describe propagation, atrophy, or pattern formation of tau and Aβ pathology (Putra et al. 2021), (Corti 2024), (Zhang et al. 2024), (Pederzoli et al. 2025), (Lee and Hao 2026). Here the primary object is instead an instantaneous kinetic rate, $R_2 = k_2 A(M) m^2$. Its maximum can move even when the maximum of fibril-mass accumulation remains fixed. Burden and activity are therefore different spatial observables.

For the same reason, the predicted kinetic hotspot should not be identified with the protofibrillar Aβ42 hotspots (Condello et al. 2015). Those plaque-local structures are shaped by microglial compaction and occur on a much smaller length scale. The present mechanism is mesoscale and emerges from monomer screening over roughly half a millimeter. A future model could couple the two by allowing microglia to modulate site accessibility, but the existing plaque-local hotspots are not direct validation of Eq. (18).

No existing experiment measures the complete set of fields required for a decisive test—tau burden, soluble Aβ, accessible-site abundance, and secondary-nucleation activity—in the same tissue and over time. Current data instead constrain individual links in the proposed causal chain: plaque-rich tissue provides a substantial monomer sink (Hong et al. 2011); seeding potency and new-plaque formation can fall while burden continues to rise (Ye et al. 2017), (Hefendehl et al. 2011), (Burgold et al. 2014); and soluble intermediates can dominate spatial transmission *in vitro* (Peter et al. 2026). These observations make the mechanism plausible without demonstrating the predicted moving hotspot.

Several tempting comparisons should therefore be resisted. Plaque-local oligomer halos and new-plaque clustering occur on scales of micrometers to tens of micrometers, well below the predicted screening length and likely reflect local seeding, transport, or microglial effects. Regional PET reorganization is also not reproduced here because the model's $\partial M/\partial t$ maximum remains at the source. These negative comparisons sharpen rather than weaken the contribution: they identify the specific phenomenon the model is trying to explain and separate it from neighboring spatial effects that require additional biology.

A study (Pooler et al. 2015) provides a limited qualitative example of spatially staged tau–amyloid coupling: increased plaque area appeared first where human tau was expressed and later in a second region after tau pathology extended there. That sequence is compatible with localized coupling, but it does not report secondary nucleation, uses tau overexpression rather than a seeding perturbation, and is confounded by neuronal loss. It therefore motivates a longitudinal co-registration experiment rather than validating the present mechanism.

The largest biological uncertainty lies at the first step of the mechanism: the strength of tau→Aβ coupling *in vivo*. The model assumes the positive heterotypic interaction measured *in vitro* (Mosconi et al. 2026). *In vivo* evidence is mixed: tau overexpression can increase plaque size or area in some models (Jackson et al. 2016), (Pooler et al. 2015), whereas reducing endogenous tau did not significantly alter plaque burden in a small 5xFAD comparison (He et al. 2018), and apparent Aβ redistribution after injection of tau extract has been attributed to Aβ contamination rather than tau itself (Nies et al. 2021). The model prediction should therefore be read conditionally: it describes the spatial consequences if a local tau→Aβ catalytic pathway operates at the disease stage being considered.

The numerical results then identify three quantitative conditions that determine whether the spatial contrast survives. First, the oligomer transport length must remain below the few-hundred-micrometer threshold above which the signal is homogenized. Second, deposits must behave as net monomer sinks over the months during which the structure forms; strong sustained release flattens the profile. Third, the apparent lead over a tau "front" depends on which contour is chosen within the diffuse Fisher–Kolmogorov tail and on domain confinement. These factors affect hotspot contrast, position, and apparent speed, but none changes the local gradient balance itself.

Growth-history-dependent site formation is different because it can reinforce the mechanism. If fibrils formed under low supersaturation carry fewer catalytic defects, then the depleted wake is penalized twice: it has less free monomer now and it retains fewer active sites per unit mass from its growth history. The Aβ40 data (Hu et al. 2026) motivate this possibility but do not calibrate it for tau-seeded Aβ42. Likewise, the amount of deposited Aβ that remains exchangeable over months is poorly constrained. These two quantities should be treated as experimental targets, not fixed model inputs.

A quantitatively predictive next model should also remove the hybrid geometry used here. Tau replication and effective spreading are taken from mouse-scale data; Aβ transport and clearance are taken from independent extracellular measurements; and the plaque-associated sink is inferred. Long segments are used only to reveal the unconfined behavior of the equations. The natural next step is therefore a network or anatomical multiscale model in which fast local Aβ equilibration is embedded in the slower evolution of tau pathology.

The most direct test would follow a spatially localized tau perturbation longitudinally while co-registering four quantities: tau, soluble Aβ, fibrillar burden or accessible-site abundance, and a readout related to secondary-nucleation activity. The model predicts that, after enough fibril ends accumulate to deplete monomer, the activity maximum should separate from the maximum tau signal. At that maximum, accessible catalytic sites should be decreasing in space while soluble Aβ is recovering, with the gradients satisfying Eq. (18). Increasing effective monomer exchange should reduce the displacement, whereas increasing the local fibril-end sink should strengthen it. Plaque burden alone cannot test this prediction.

**Table 3. Testable spatial predictions of the model.**

| Prediction | Mechanism in the model | Measurement needed |
|---|---|---|
| The secondary-nucleation maximum can be displaced from the maximum tau burden. | High tau first raises $P$ and accessible catalytic sites, then lowers $m$ through elongation. | Co-registered tau, soluble Aβ, plaque burden or active-site abundance, and secondary-nucleation readouts. |
| Hotspot position satisfies a gradient balance. | At $x_h$, $\partial_x \ln A = -2\,\partial_x \ln m$ [Eq. (18)]. | Spatial gradients of accessible catalytic sites and soluble Aβ. |
| Faster monomer redistribution suppresses hotspot displacement. | Lower $\mathrm{Da}_D$ means that diffusion smooths monomer gradients faster than elongation can sustain them. | Perturb extracellular transport/clearance and track hotspot position. |
| In the Fisher–Kolmogorov implementation, the nucleation maximum can lead a chosen tau-concentration contour once confinement is removed. | The maximum follows Eq. (18) and can lie in the diffuse leading tail of $T$; the numerical lead depends on the contour definition. | Longitudinal co-registration of tau concentration contours and new-deposit or secondary-nucleation activity over a field large enough to keep the maximum interior. |
| Plaque burden alone can miss the most active region. | $R_2$ depends on $A(M)m^2$, so neither fibril mass nor accessible catalytic sites alone determines it. | Measure active sites and free monomer in addition to fibril mass. |

## 7. Conclusions: the most active region need not carry the greatest burden

If a local tau→Aβ catalytic pathway operates, adding space to active-site-resolved aggregation kinetics produces a counterintuitive but simple prediction: the region with the greatest tau burden or fibril mass need not have the greatest instantaneous Aβ secondary-nucleation activity. Tau first creates fibrils and accessible catalytic sites; continued fibril growth then consumes the monomer required to use those sites. An interior maximum appears where the logarithmic decrease in accessible catalytic sites is balanced by twice the logarithmic recovery of monomer [Eq. (18)]. $\mathrm{Da}_{\mathrm{EC}}$ determines whether elongation can deplete monomer locally, whereas $\mathrm{Da}_D$ and the screening length determine whether diffusion can erase that depletion. Independent transport and turnover data place the relevant screening length near 0.44–0.54 mm, making the predicted displacement a mesoscale tissue phenomenon rather than a single-plaque effect.

The numerical calculations also define the limits of that prediction. Faster monomer redistribution, long-range oligomer transport, or sustained reversible release can flatten the spatial profile; growth-history-dependent catalytic-site density can strengthen it. The apparent motion of the maximum relative to a Fisher–Kolmogorov tau contour is not universal and should not be interpreted as Aβ activity propagating through tau-free tissue. The more robust experimental target is the spatial separation itself: tau burden, fibril-mass growth, and secondary-nucleation activity need not peak at the same place. Testing that separation requires co-registered measurements of tau, soluble Aβ, catalytic-site abundance or activity, and longitudinal new-deposit formation. The key biological uncertainty remains whether the positive tau→Aβ coupling measured *in vitro* is strong enough to operate *in vivo*.

### Data and Code Availability Statement

No experimental data were generated in this study. All numerical results were obtained from Eqs. (11)–(15), and from the quasi-steady reduction of Eq. (16) where stated, with the parameters in Table 1 and the dimensional mapping of Section 2.6, using the MATLAB implementation described in Section 4. The solver, the diagnostic routines, the parameter sets, a driver that reports every numerical result of Sections 4 and 5, and a script that regenerates Figs. 2–9 accompany the manuscript and are archived at Zenodo (Kuznetsov 2026b). Running that driver reproduces the tabulated hotspot positions, the detachment and source-maximum times, the sensitivity to monomer redistribution at every value of $D_m$, and the grid-convergence results of Section 4, subject to the peak-location convention stated there.

### Funding

The author acknowledges support provided by the National Science Foundation (grant DMS-2451660) and the Alexander von Humboldt Foundation through the Humboldt Research Award.

### Competing interests

The author declares no competing interests.

### Author contributions

A.V.K. conceived the study, formulated and analyzed the model, wrote the software, performed the calculations, prepared the figures, and wrote the manuscript.